**Extrinsic Size Effect in Piezoresponse Force Microscopy of Thin Films**

Anna N. Morozovska,[1] Sergei V. Svechnikov,[1] Eugene A. Eliseev,[*,2] and Sergei V. Kalinin[†,3]

[1]Institute of Semiconductor Physics, National Academy of Science of Ukraine,

45, pr. Nauki, 03028 Kiev, Ukraine

[2]Institute for Problems of Materials Science, National Academy of Science of Ukraine,

3, Krjijanovskogo, 03142 Kiev, Ukraine

[3]Materials Science and Technology Division and Center for Nanophase Materials Science,

Oak Ridge National Laboratory, Oak Ridge, TN 37831

The extrinsic size effect in Piezoresponse Force Microscopy of ferroelectric and piezoelectric thin films on non-polar dielectric substrate is analyzed. Analytical expressions for effective piezoresponse, object transfer function components, and Rayleigh two-point resolution are obtained. These results can be broadly applied for effective piezoelectric response calculations in thin piezoelectric and ferroelectric films as well as surface polar layers e.g. in organic materials and bio-polymers. In particular, the effective piezoresponse strongly decreases with film thickness whereas the sharpness of domain stripes image increases due to the object transfer function spectrum broadening.

---

[*] Corresponding author, eliseev@i.com.ua

[†] Corresponding author, sergei2@ornl.gov

Piezoelectric and dielectric properties of thin films, including ferroelectric properties, piezoelectric coupling coefficients, and domain structures, can be strongly modified compared to the bulk. This intrinsic ferroelectric size effect in thin films and heterostructures presents considerable interest for applications such as non-volatile ferroelectric memories,[1,2] ferroelectric data storage,[3,4] and resistive memories based on ferroelectric and multiferroic tunneling barriers.[5] At the same time, surfaces of non-piezoelectric materials can develop built-in dipole moment due to the inversion symmetry breaking and thus possess surface piezo- and flexoelectricity.[6, 7, 8] For example, a dipole layer on the surface such as an ordered water layer is expected to be piezoelectric, whereas a liquid-like water layer will be non-piezoelectric. This electromechanical coupling is directly linked to the phenomena such as sonoluminiscence, cavitation, and triboelectricity. Hence, the knowledge of local piezoelectric response of thin films and surface layers as well as their domain structures is a key component for the verification of theoretical models, design of functional nanomaterials with predetermined properties and also application in various devices (such as ferroelectric micro and nanocapacitors, sensors and actuators, etc.)

The development of Piezoresponse Force Microscopy (PFM)[9, 10] and its spectroscopic variants has allowed 2D mapping of domain structures and switching behavior in ferroelectric thin films. One of the issues that have recently attracted much attention is the evolution of the piezoelectric and ferroelectric properties with the film thickness.[11,12,13] Many of these studies utilize PFM as an effective alternative to double-interferometer based capacitor measurements.[14] Furthermore, PFM is a natural technique for the low-dimensional structures and ultrathin films, in which spatial inhomogeneity and pinholes effectively preclude capacitor-based measurements. However, in the tip-electrode PFM experiment, the electric field produced by the system is not uniform, potentially affecting the absolute value of PFM signal. Notably, the use of simple 1D approximation, equivalent to plane capacitor geometry, does not capture this effect. Here we derive analytical results for the effective PFM response of tetragonal piezoelectric film (or surface layer) on semi-infinite substrate (or bulk) and analyze thickness effect on resolution. These results are thus relevant both for the quantitative interpretation of PFM data on ultrathin films and development of efficient strategies for probe-based high-density data storage.

Recently a universal approach based on the decoupled theory[15,16] has been developed to derive analytical expressions for PFM response on semi-infinite materials.[17] Assuming that piezoelectric coupling is longitudinally-homogeneous within the penetration depth of the probe electric field or within the thickness of ultrathin film, the surface piezoresponse below the tip [i.e. the surface displacement $u_i(\mathbf{x}, \mathbf{y})$ in the point $x_1 = x_2 = 0$] is given by the convolution of an *ideal image* $d_{klj}$ with the resolution function components $R_{ijkl}(\xi_1, \xi_2)$ [18, 19]:

$$u_i(\mathbf{y}) = \int_{-\infty}^{\infty} d\xi_1 \int_{-\infty}^{\infty} d\xi_2 \, R_{ijkl}(\xi_1, \xi_2) d_{lkj}(y_1 - \xi_1, y_2 - \xi_2),$$

$$R_{ijkl}(\xi_1, \xi_2) = \int_0^h d\xi_3 c_{kjmn} \frac{\partial G_{im}(-\xi_1, -\xi_2, \xi_3)}{\partial \xi_n} E_l(\xi_1, \xi_2, \xi_3). \quad (1)$$

Here $h$ is the film thickness, $\mathbf{x} = (x_1, x_2, x_3)$ is position with respect to the tip apex, and $\mathbf{y} = (y_1, y_2)$ is the tip position in the sample coordinate system $\mathbf{y}$ (Fig. 1). Coefficients $d_{ijk}(y_1, y_2)$ are the film piezoelectric tensor, $c_{kjmn}$ are elastic stiffness tensor components. The mechanical properties are described by the elastic Green's function $G_{3j}(\mathbf{x} - \boldsymbol{\xi})$ that depends on Young modulus and Poisson ratios $\nu$ (typically $\nu \sim 0.25 – 0.35$). Here we use the Green's function of a semi-infinite elastically isotropic medium, i.e., assume that the elastic properties of film and substrate are close.

The *ac* electric field distribution produced by the probe is $E_k(\mathbf{x}) = -\partial\varphi(\mathbf{x})/\partial x_k$, where $\varphi$ is *ac* component of electrostatic potential determined by the tip geometry. Here we derive the PFM response components for effective point charge model of charge $Q$ located at the distance, $d$, from the surface. Below we illustrate that the point charge model in certain cases can be used to describe PFM response for the flattened tip-surface contact, corresponding to experimentally observed case.[20]

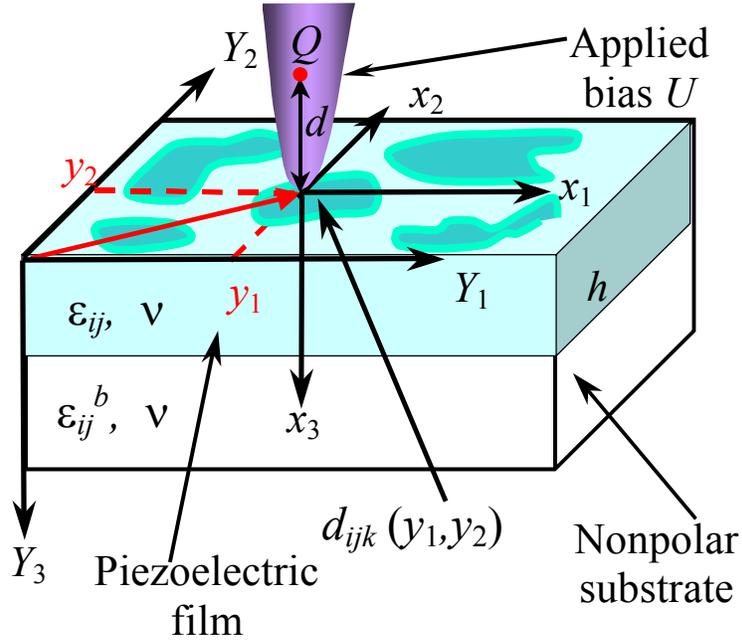

**Fig. 1.** (Color online) Coordinate systems and considered structure in PFM experiment.

For the systems with one boundary (i.e., semi-infinite material) rigorous potential structure can be obtained using image-charge series for $\varphi$ (see e.g., Ref. 21). For the multilayer system "ambient - film - substrate" with different dielectric constants, the series solution is extremely complex (see Ref. 22), necessitating the development of point-charge like model. For PFM, the ratio $Q/d$ is selected to give applied potential $U$ in the point of tip-sample contact. The solution is found as a function of reduced thickness $h/d$ as:[23]

$$\frac{Q(h,d)}{d} = 2\pi\varepsilon_0 (\varepsilon_e + \kappa) U \cdot \psi(h,d),$$

$$\psi(h,d) \approx \left(1 + \left(\frac{\kappa_b + \varepsilon_e}{\kappa - \kappa_b} + \frac{h}{\gamma d}\frac{\varepsilon_e - \kappa}{\kappa}\ln^{-1}\left(1 - \frac{\kappa_b - \kappa}{\kappa_b + \kappa}\cdot\frac{\varepsilon_e - \kappa}{\varepsilon_e + \kappa}\right)\right)^{-1}\right)^{-1}. \quad (2)$$

Here $\gamma = \sqrt{\varepsilon_{33}/\varepsilon_{11}}$ is the dielectric anisotropy factor of the film, $\varepsilon_e$ is the dielectric constant of the ambient, $\kappa = \sqrt{\varepsilon_{33}\varepsilon_{11}}$ and $\kappa_b = \sqrt{\varepsilon_{33}^b \varepsilon_{11}^b}$ are the effective dielectric permittivity values of the film and the substrate. In order to derive the quantities of $Q$ and $d$ separately, Eq. (2) should be complemented by condition that can be chosen to represent (a) tip curvature in the point of contact, (b) total tip charge, or (c) electrostatic field produced by the contact part of the probe. These three possible point charge models are described below.

In the framework of the effective point charge model[24] the isopotential surface curvature reproduces the tip curvature in the point of contact. This model is appropriate for electric field description in the immediate vicinity of the tip-surface junction, relevant for e.g. modeling nucleation processes. For the film thickness $h \geq d$, the effective point charge model gives $d \approx \varepsilon_e R_0/\kappa$ for the spherical tip with curvature $R_0$.

The field structure in the most part of piezoresponse volume,[22] can be represented by the point charge model in which the effective charge value $Q$ is equal to the product of tip capacitance on applied voltage. For the film thickness $h \geq d$, it gives the effective separation as $d \approx 2\varepsilon_e R_0 \ln((\varepsilon_e + \kappa)/2\varepsilon_e)/(\kappa - \varepsilon_e)$ for the spherical tip with curvature $R_0$.

Finally, for the conductive disk of radius $R_0$ representing contact area, Eq. (2) can be used with $d \approx 2R_0/\pi$. The proposed model satisfies the condition (2), thus it should be clearly distinguished from the conventional capacitance approximation (with $d = R_0$) that describes electric field far from the tip only.

For the laterally homogeneous film (piezoelectric tensor is $d_{ijk}$ within the film and 0 in the substrate), the effective vertical piezoresponse $d_{33}^{eff} = u_3(\mathbf{y}=0)/U$ is derived as

$$d_{33}^{eff}(h) = -\psi(h)\left(W_{313}(h)d_{31} + W_{333}(h)d_{33} + W_{351}(h)d_{15}\right), \quad (3)$$

where $W_{313}(h) \approx (1+2\nu)W_{313}^{\nu}(h) - W_{313}^{0}(h)$, and

$$W_{313}^{\nu}(h) \approx \left(1 + \frac{\kappa_b - \kappa}{\kappa_b + \kappa}\frac{\gamma d}{\gamma d + 2h}\right)\frac{h}{(\gamma d + (1+\gamma)h)}, \quad (4a)$$

$$W_{313}^{0} \approx \left(1 + \frac{\kappa_b - \kappa}{\kappa_b + \kappa}\frac{\gamma d}{\gamma d + 2h}\right)\frac{\gamma h^2}{(\gamma d + (1+\gamma)h)^2}, \quad (4b)$$

$$W_{333}(h) \approx \left(1 + \frac{\kappa_b - \kappa}{\kappa_b + \kappa}\frac{\gamma d}{\gamma d + 2h}\right)\frac{h(h + (d+2h)\gamma)}{(\gamma d + (1+\gamma)h)^2}, \quad (4c)$$

$$W_{351}(h) \approx \left(1 - \frac{\kappa_b - \kappa}{\kappa_b + \kappa}\frac{\gamma d}{\gamma d + 2h}\right)\frac{\gamma^2 h^2}{(\gamma d + (1+\gamma)h)^2}. \quad (4d)$$

Function $\psi(h)$ is given by Eq. (2). Eqs. (4) corresponds to the first two terms (point charge + its first image) of the full image charge series given in Ref. 22. The accuracy of Eqs. (4) is better than 5% for $\varepsilon_e < \kappa < \kappa_b$, even for significantly different dielectric permittivity $\kappa$ and $\kappa_b$, since the series is alternating and converges very fast. For $\kappa > \kappa_b > \varepsilon_e$, the series of constant signs converges more slowly, and accuracy better than 10% can be achieved only for $|(\kappa_b - \kappa)/(\kappa_b + \kappa)| \leq 0.5$.

Under the assumption $\left|(\kappa_b - \kappa)/(\kappa_b + \kappa)\right| \leq 0.2$, the piezoresponse, $d_{33}^{eff}$, of the film is:

$$d_{33}^{eff}(h) \approx -\frac{\gamma^2 h^2 d_{15}}{(h+\gamma(d+h))^2} - \frac{h(h+(d+2h)\gamma)}{(h+\gamma(d+h))^2}d_{33} - \left(\frac{(1+2\nu)h}{h+\gamma(d+h)} - \frac{\gamma h^2}{(h+\gamma(d+h))^2}\right)d_{31} \quad (5)$$

A number of non-trivial consequences can be derived from Eq. (5) for the PFM contrast of ultrathin films ($h \ll d$). The effective piezoresponse scales linearly with film thickness, $h$, as $d_{33}^{eff} \approx -(d_{33} + (1+2\nu)d_{31})h/\gamma d$. Remarkably, ultrathin film response $d_{33}^{eff}$ is proportional to $d_{33}$ and $d_{31}$, but it is independent on $d_{15}$ (rigorously, the contribution of $d_{15}$ scales as a square on the film thickness). It becomes clear since the components proportional to $d_{33}$ and $d_{31}$ originate from the coupling with longitudinal tip electric field $E_3(\rho,0)$ that is maximal just below the tip, whereas components proportional to $d_{15}$ couples with transverse fields $E_{1,2}(\rho,0)$ that is zero directly below the tip.

The extrinsic size effect on effective piezoresponse $d_{33}^{eff}(h)$ is illustrated in Fig. 2 for ferroelectric PbTiO$_3$ deposited on the non-piezoelectric substrates with different dielectric constants $\kappa_b$ (compare the curves form with experimental results for PZT/SRO thin films of Fig. 5 in Ref.[11]. In this case, finite screening length of SRO will give rise to qualitatively similar behavior).

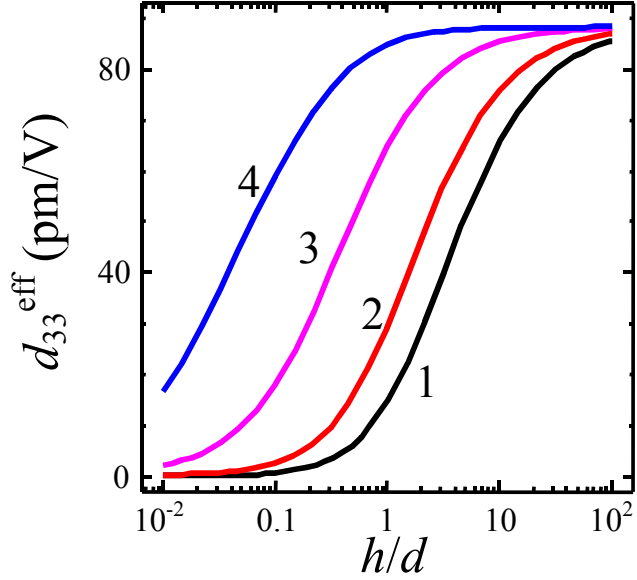

**Fig. 2.** (Color online) Effective piezoelectric response $d_{33}^{eff}(h)$ of PbTiO$_3$ ($\kappa = 121$, $\gamma = 0.87$) film of thickness $h$ capped on substrate with the same elastic properties ($\nu = 0.3$) and different dielectric constants $\kappa_b = 3; 30; 300; 3\times10^3$ (curves 1, 2, 3, 4 respectively); $\varepsilon_e = 1$.

Note that response is thickness-dependent for $h/d \leq 10^2$, i.e., film thickness should exceed the effective tip radius by 1-2 orders of magnitude for response saturation. For instance, for tip radius of 10 nm the PbTiO$_3$ film thickness should be 25-250 nm for response to saturate to 90% of bulk value for $\kappa_b = 3 - 300$. This *extrinsic* size effect should be distinguished from the *intrinsic* ones in thin films related to thickness dependence of the polarization and piezoelectric coefficients.

The effective tip radius can be determined from the PFM imaging on suitably chosen calibration standard in the form of thick film or single crystal of the same material. However, it cannot be determined self-consistently due to $h$-dependence of resolution, as discussed in detail below.

This analysis can be extended for the case when the film is inhomogeneous in the transverse directions $\{y_1, y_2\}$ using phenomenological resolution function theory.[18,19] For linear imaging, experimentally measured image $d_{ijk}^{eff}(\mathbf{x})$ is the convolution of the ideal image $d_{ijk}(\mathbf{x})$ with the *resolution function* $R_{ijk}(\mathbf{y})$. This relationship can be conveniently represented in the Fourier domain using tensorial *object transfer function* (OTF) components $\widetilde{R}_{ijk}(\mathbf{q})$ relating ideal image $\widetilde{d}_{ijk}(\mathbf{q})$ and experimentally measured image $\widetilde{d}_{ijk}^{eff}(\mathbf{q})$. The Fourier transform of film vertical piezoresponse $\widetilde{d}_{33}^{eff}(\mathbf{q})$ over transverse coordinates $\{y_1, y_2\}$ is

$$\widetilde{d}_{33}^{eff}(\mathbf{q}) = -\widetilde{R}_{313}(\mathbf{q})\widetilde{d}_{31}(\mathbf{q}) - \widetilde{R}_{333}(\mathbf{q})\widetilde{d}_{33}(\mathbf{q}) - \widetilde{R}_{351}(\mathbf{q})\widetilde{d}_{15}(\mathbf{q}). \tag{6}$$

Under the condition $|(\kappa_b - \kappa)/(\kappa_b + \kappa)| \leq 0.5$ we derived Pade approximation for OTF[22]:

$$\widetilde{R}_{313}(q) \approx (1+2\nu)\frac{W_{313}^{\nu}(h) \cdot \psi_i(h)}{1 + \gamma W_{313}^{\nu}(h)qd} - \frac{W_{313}^{0}(h) \cdot \psi_i(h)}{1 + \gamma W_{313}^{0}(h)qd}, \tag{7a}$$

$$\widetilde{R}_{333}(q) \approx \frac{2W_{333}(h) \cdot \psi_i(h)}{2 + \gamma W_{333}(h)qd}, \quad \widetilde{R}_{351}(q) \approx \frac{6W_{351}(h) \cdot \psi_i(h)}{6 + \gamma W_{351}(h)(qd)^3}, \tag{7b}$$

where $q = \sqrt{q_1^2 + q_2^2}$ and $W_{3jk}(h)$ are listed after the Eqs. (4).

As an example, we calculate the effective piezoresponse of stripe domain structure with polarization $\pm P_S$ and period $a$ (see Fig. 3).

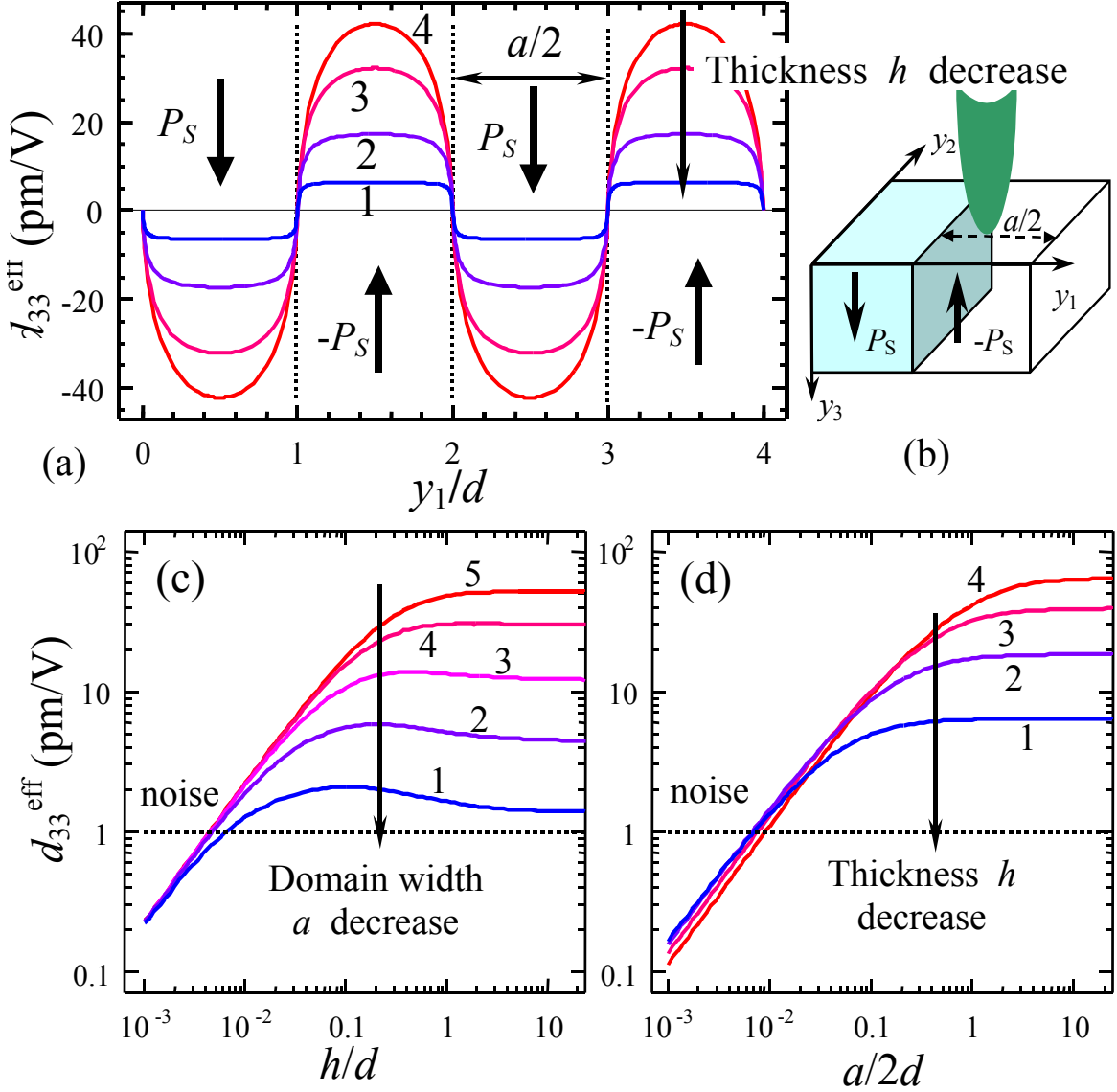

**Fig. 3.** (Color online) (a) PFM profile of periodic stripe domain structure (b) in PbTiO$_3$ film on SrTiO$_3$ substrate for different film thickness $h/d = 0.03, 0.1, 0.3, 1$ (curves 1, 2, 3, 4). (c) Maximal piezoresponse vs. film thickness for different stripe period $a/d = 0.03, 0.1, 0.3, 1, 3$ (curves 1, 2, 3, 4, 5). (d) Dependence of response on stripe period for different film thicknesses $h/d = 0.03, 0.1, 0.3, 1$ (curves 1, 2, 3, 4).

Using the rectangular wave Fourier series for $\tilde{d}_{ijk}(\mathbf{q})$ (see e.g., Ref. 19) and Eqs. (6-7), the vertical piezoresponse is

$$d_{33}^{eff}(y_1) = -\sum_{n=0}^{\infty} \frac{4\sin(q_n y_1)}{(2n+1)\pi} \left( \tilde{R}_{313}(q_n) d_{31} + \tilde{R}_{333}(q_n) d_{33} + \tilde{R}_{351}(q_n) d_{15} \right), \quad (8)$$

where the summation is performed over wave vectors $q_n = 2\pi(2n+1)/a$. PFM profile of periodic stripe domain structure in PbTiO$_3$ film on SrTiO$_3$ substrate for different film thickness, $h/d$, is shown in Figs. 3a,b. It is clear that under the film thickness decrease the profile $d_{33}^{eff}(y_1)$ acquires rectangular shape, more close to the ideal image. Note, that the sharpness of domain stripes image increases due to the object transfer function spectrum broadening.[22] At the same time, the signal strength decreases for smaller film thicknesses, making the relative noise level higher (see Figs. 3c,d).

Maximal information limit (i.e. domain size such that domain is still observable above noise level) of the domain stripes calculated from the equation $d_{33}^{eff}(a,h) = n$ is shown in Fig.4 for different noise level $n$. Corresponding 3D plot of maximal piezoresponse $d_{33}^{eff}(a,h)$ is shown in the inset. The periodic domain structure can be unambiguously resolved with maximal value exceeding some noise level in the region above the given curve related to this noise level. Below this limit the structure image will be masked by the noise. The critical film thickness $h_{cr}(n)$ corresponds to the case when even the piezoresponse of homogeneous layer becomes smaller than the noise level $n$. It is non-trivial that the finest domain structure ($a/2d < 10^{-2}$) could be resolved in ultra-thin films ($10^{-2} < h/d < 1$). It should be noted that the minimal resolved stripe period is only weakly dependent on the film thickness; this dependence becomes abrupt only in the vicinity of the critical thickness $h_{cr}(n)$, corresponding to PFM amplitude becoming smaller then the noise level. Hence, experimentally domains will be observable continuously with thickness decrease at reduced amplitude but increased sharpness, and eventually rapidly "fade off" once noise level is achieved. This behavior suggests the optimal strategy for development of ferroelectric thin film data storage, in which films with thickness slightly above critical will provide highest read-out density.

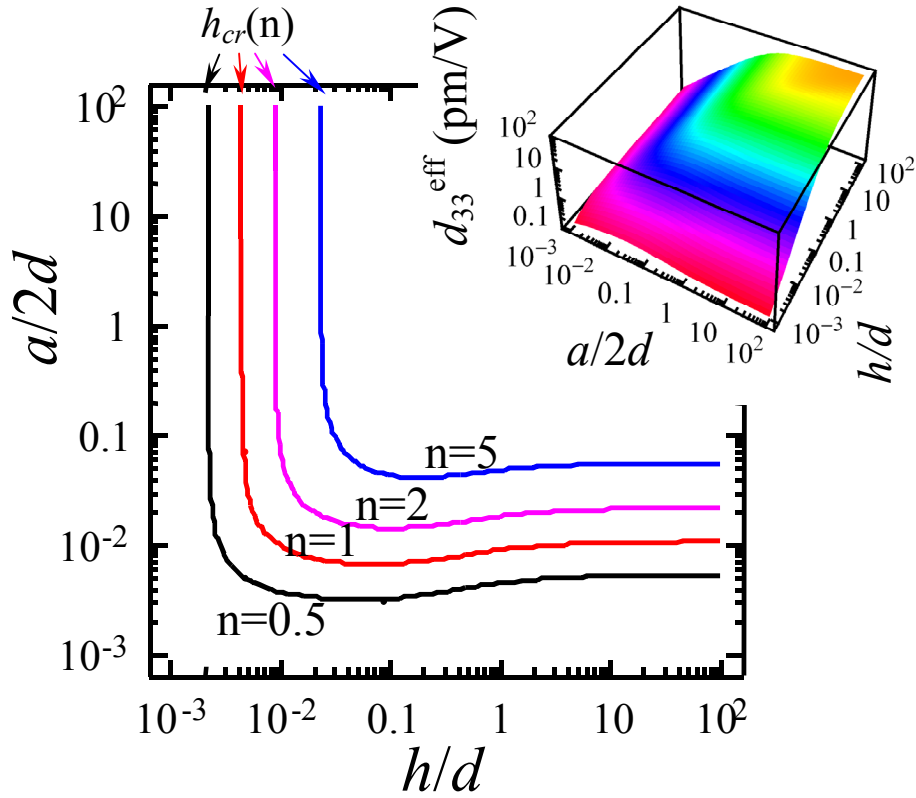

**Fig. 4.** Information limit for maximal piezoresponse $d_{33}^{eff}(a,h)$ of the domain stripes with period $a$ in the PbTiO$_3$ film of thickness $h$ on SrTiO$_3$ substrate for different noise level $n$ (figures near the curves in pm/V). 3D plot of maximal piezoresponse $d_{33}^{eff}(a,h)$ is shown in the inset.

To summarize, we derive the analytical description of *extrinsic* size effect in PFM imaging of thin piezoelectric films. The effective piezoresponse $d_{33}^{eff}$ exhibits decrease with thickness for $h < 10d$, corresponding to 100 nm films and below. This effect is especially pronounced for substrates with low dielectric constants. These considerations should be taken into account in any quantitative PFM experiments on thin films or surface polar layers, in which thickness dependence of PFM signal is used as an indicator of size-dependent piezoelectric activity. For laterally inhomogeneous films, the resolution and transfer function components are derived as a function of thickness. Furthermore, this analysis is applicable for calculation of effective piezoresponse of surface layers in non-polar and polar materials.

Research supported in part (SVK) by Division of Materials Science and Engineering, Oak Ridge National Laboratory, managed by UT-Battelle, LLC, for the U.S. Department of Energy under Contract DE-AC05-00OR22725.

References

[1] J. Scott, *Ferroelectric Memories* (Springer Verlag, Berlin, 2000).

[2] Rainer Waser (Ed.), *Nanoelectronics and Information Technology*, Wiley-VCH (2003)

[3] T. Tybell, C. H. Ahn, and J.-M. Triscone, Appl. Phys. Lett. **72**, 1454 (1998)

[4] T. Tybell, P. Paruch, T. Giamarchi, and J.-M. Triscone, Phys. Rev. Lett. **89**, 097601 (2002).

[5] E.Y. Tsymbal and H. Kohlstedt, Science **313**, 181 (2006).

[6] A.K. Tagantsev, Phys. Rev. **B 34**, 5883 (1986).

[7] M. D. Glinchuk and A. N. Morozovska. J. Phys.: Condens. Matter **16**, 3517 (2004).

[8] A. M. Bratkovsky, and A. P. Levanyuk, Phys. Rev. Lett. **94**, 107601 (2005).

[9] S.V. Kalinin, A. Rar, and S. Jesse, IEEE TUFFC **53**, 2226 (2006).

[10] A. Gruverman and A. Kholkin, Rep. Prog. Phys. **69**, 2443 (2006).

[11] V. Nagarajan, J. Junquera, J.Q. He, C.L. Jia, R. Waser, K. Lee, Y.K. Kim, S. Baik, T. Zhao, R. Ramesh, Ph. Ghosez, and K.M. Rabe, J. Appl. Phys. **100**, 051609 (2006).

[12] C. Lichtensteiger, J.-M. Triscone, Javier Junquera and Ph. Ghosez, Phys. Rev. Lett. **94,** 047603 (2005).

[13] C.L. Jia, V. Nagarajan, J.Q. He, L. Houben, T. Zhao, R. Ramesh, K. Urban, and R. Waser, Nature Mat. **6**, 64 (2007).

[14] A. L. Kholkin, Ch. Wütchrich, D. V. Taylor, and N. Setter, Rev. Sci. Instrum. **67**, 1935 (1996)

[15] F. Felten, G.A. Schneider, J.M. Saldaña, and S.V. Kalinin, J. Appl. Phys. **96**, 563 (2004).

[16] D.A. Scrymgeour and V. Gopalan, Phys. Rev. **B 72**, 024103 (2005).

[17] S.V. Kalinin, E.A. Eliseev, and A.N. Morozovska, Appl. Phys. Lett. **88**, 232904 (2006).

[18] S.V. Kalinin, S. Jesse, J. Shin, A.P. Baddorf, H.N. Lee, A. Borisevich, and S.J. Pennycook, Nanotechnology **17**, 3400 (2006).

[19] A.N. Morozovska, S.L. Bravina, E.A. Eliseev, and S.V. Kalinin. Accepted to Phys. Rev. **B**, cond-mat/0608289 (unpublished) (2006).

[20] V. Gopalan and L. Tian, private communications

[21] Catalin Harnagea, Ph.D. thesis, Martin-Luther-Universität Halle-Wittenberg, 2001.

[22] A.N. Morozovska, E.A. Eliseev, E. Karapetian and S.V. Kalinin (unpublished)

[23] Note that universal length-scale exists only to linear cases such as imaging, and its applicability for switching is not guaranteed.

[24] A. N. Morozovska, E.A. Eliseev, and S.V. Kalinin. Appl. Phys. Lett. **89**, 192901 (2006).